\documentclass[showpacs, twocolumn]{revtex4-1}
\usepackage{graphicx}
\usepackage{amsmath}
\usepackage{appendix}

\begin{document}

\title{Theory of excitations of dipolar Bose-Einstein condensate at finite temperature}

\author{Abdel\^{a}ali Boudjem\^{a}a}

\affiliation{Department of Physics, Faculty of Sciences, Hassiba Benbouali University of Chlef P.O. Box 151, 02000, Ouled Fares, Chlef, Algeria.}


\begin{abstract}

We present  a systematic study of dilute three-dimensional dipolar Bose gas employing a finite temperature perturbation theory (beyond the mean field). 
We analyze in particular the behavior of the anomalous density, we find that this quantity has a finite value in the limit of weak interactions at both zero and finite temperatures.
We show that the presence of the dipole-dipole interaction (DDI) enhances fluctuations, the second order correlation function
 and thermodynamic quantities such as the chemical potential, the ground state energy, the compressibility and the superfluid fraction.  
We identify the validity criterion of the small parameter of the theory for Bose-condensed dipolar gases.

\end{abstract}


\maketitle

\section{Introduction}

The recent experimental realization of Bose-Einstein condensate (BEC) of ${}^{52}$Cr \cite{Pfau}, ${}^{164}$Dy \cite{ming},  ${}^{168}$Er \cite{erbium} 
and more recently with a degenerate Fermi gas of ${}^{161}$Dy \cite{lu} atoms with large magnetic dipolar interaction  (6 $\mu_B$, $10\mu_B$ and $7\mu_B$, respectively) 
has opened fascinating prospects for the observation of novel quantum phases and many-body phenomena\cite{Baranov,Pfau,Carr,Pupillo2012}. 
Polar molecules which have much larger electric dipole moments than those of the atomic gases have been also produced in their ground rovibrational state \cite {Aik,kk}.
The most important feature of these systems is that the atoms interact via a DDI that is both long ranged and anisotropic.
The anisotropy introduced by the dipolar interactions manifests in the expansion dynamics \cite{Pfau}, the excitation spectrum \cite {bism}, superfluid properties \cite{Tic, Odell}, 
solitons and soliton-molecule\cite{Tik, santos2, Adhi}. 
Additionally, the DDI is partially attractive and exhibits a roton-maxon structure in the spectrum \cite {santos1} and enhances fluctuations \cite {abdougora, Blak2}.
On the other hand, the long range character of the dipolar interaction leads to scattering properties that are radically 
different from those found on the usual short-ranged potentials of quantum gases and therefore, all of the higher-order partial waves 
contribute equally to the scattering at low energy \cite {Baranov}.

The majority of theoretical investigation of BEC with DDI has often been focused on zero temperature case described by either the Gross-Pitaevskii equation or the standard Bogoliubov approximation \cite {Santos, Eberlein, santos1, dell, lime}. These works have studied in particular, excitations, ground-state properties and the stability of dipolar BECs. 
In contrast, few are the attempts directed towards the finite temperature behavior of dipolar Bose gases, 
for instance we can quote path-integral Monte Carlo simulations \cite{Nho, Filin} 
and mean field Hartree-Fock-Bogoliubov  (HFB) theory  within numerical calculations\cite{Ron, Blak, Hut, Blak1}.

The above theories, although being satisfactory, they suffer from several drawbacks. 
First, the HFB approximation is not able in principle to describe the condensed state with broken gauge symmetry, 
since the breaking of gauge symmetry is a necessary and sufficient condition for BEC \cite {Lieb}. 
Also, the HFB approximation leads to an unphysical gap in the excitation spectrum, which causes a violation of
the Hugenholtz-Pines (HP) theorem \cite {HP}. The standard Bogoliubov approximation \cite {Bog} by its construction 
is applicable only at zero or at very low temperatures, where the Bose-condensed fraction is dominant. 
Furthermore, in many approaches, the anomalous density is neglected under the claim that it is an
unmeasured quantity, as well as its contribution being very small compared to the noncondensed density. 
In fact, it is not obvious that it is consistent to calculate the noncondensed density self-consistently and
ignore the anomalous density, since the lowest order interaction contributions to both quantities are found to be of the same order in three- and 
two-dimensional Bose gas \cite {Griffin, Burnet, Yuk, boudj2010, boudj2011, boudj2012} with contact interactions. 
Moreover, it has been proved that this quantity plays a crucial role on the stability of the system.
By definition, the anomalous average arises of the symmetry breaking assumption \cite{Yuk, boudj2012} and
it quantifies the correlations between pairs of condensed atoms with pairs of noncondensed atoms. 

In this paper we present a full self-consistent theory to study the properties of three-dimensional homogeneous dipolar Bose gas at finite temperature. 
This method which based on Beliaev's higher order (finite temperature) of perturbation theory being gapless and conserving.
For homogeneous gases, Beliaev \cite{Bel} was developed the theory beyond the mean field approach by constructing 
the zero-temperature diagram technique which allows one to find corrections to the energies of Bogoliubov excitations, 
proportional to $ \sqrt{ n_c a^3 }$, where $n_c$ is the condensate density. 
For BECs with contact interactions, Beliaev's work was extended by several authors \cite{pop, Fed, Griffin} at finite temperatures.

The rest of the paper is organized as follows. In section\ref {model}, we review the main steps, which the model is based on. 
In section \ref{zeroT}, we study the quantum fluctuations and their effects on the thermodynamics of the system. 
We examine in particular the behavior of the anomalous density and its effects on the second order correlation function at zero temperature.
Moreover, we show that the DDIs enhance the condensate depletion, the anomalous density and thermodynamic quantities such as 
the chemical potential, the ground state energy and the compressibility. The universal small parameter of the theory is also established . 
In section \ref{finiteT}, we extend our results to the finite temperature case. 
Finally, our conclusions are drawn in section\ref{conc}.

\section{The model} 
\label {model}
We consider a dilute Bose gas with $N$ dipoles aligned along the $z$ axis, in this case the interaction potential has a contact component related to the s-wave scattering length $a$ as
$V_c(\vec r)=g\delta (\vec r)=(4\pi \hbar^2 a/m)\delta (\vec r)$, and the dipole-dipole component which reads  
\begin{equation}\label{dd}
V_d(\vec r) =\frac{C_{dd}}{4\pi}\frac{1-3\cos^2\theta}{r^3},
\end{equation}
where the coupling constant $C_{dd} $ is $M_0 M^2$ for particles having a permanent magnetic dipole moment $M$ ($M_0$ is the magnetic permeability
in vacuum) and $d^2/\epsilon_0$ for particles having a permanent electric dipole $d$ ($\epsilon_0 $ is the permittivity of vacuum),
$m$ is the particle mass, and $\theta$ is the angle between the relative position of the particles $\vec r$ and the direction of the dipole. \\
The characteristic dipole-dipole distance can be defined as $r_*=m C_{dd}/4\pi \hbar^2$ \cite {abdougora}. For most polar molecules $r_*$ ranges from 10 to $10^4$ \AA. 

In the ultracold limit where the particle momenta satisfy the inequality $kr_*\ll1$, the scattering amplitude is given by \cite{Baranov}
\begin{equation}\label{scam1}
 f(\vec k)=g[1+\epsilon_{dd} (3\cos^2\theta{_k}-1)], 
\end{equation}
where the vector $\vec k$ represents the momentum transfer imparted by the collision, and 
$\epsilon_{dd}=C_{dd}/3g $ is the dimensionless relative strength which describes the interplay between the DDI and short-range interactions.
The expression (\ref{scam1}) can be obtained also using the Fourier transfrom\cite {Pfau, Baranov, Carr}.
Employing this result in the second quantized Hamiltonian, we obtain in the uniform case
\begin{eqnarray}\label{he3}
\!\!\!\!\hat H\!\!=\!\!\sum_{\vec k}\!\frac{\hbar^2k^2}{2m}\hat a^\dagger_{\vec k}\hat a_{\vec k}\!+\!\frac{1}{2V}\!\!\sum_{\vec k,\vec q,\vec p}\!\! f (\vec p)
\hat a^\dagger_{\vec k+\vec p} \hat a^\dagger_{\vec q-\vec p}\hat a_{\vec q} \hat a_{\vec k},
\end{eqnarray}
where $V$ is a quantization volume, and $\hat a_{\vec k}^\dagger$, $\hat a_{\vec k}$ are
the creation and annihilation operators of particles.\\
In Hamiltonian (\ref{he3}), the first term in the single-particle part corresponds to the kinetic energy of particles 
and the second term describes the two-body interaction Hamiltonian of the dipolar force.\\
Assuming the weakly interacting regime where  $r_*\ll \xi_c$ with $\xi_c=\hbar/\sqrt{mgn_c}$ being the corrected healing length,
we may use the Bogoliubov approach up to the fourth order of perturbation theory.
Employing the canonical Bogoliubov transformations: 
\begin{equation}\label {trans}
\hat a^\dagger_{\vec k}= u_k \hat b^\dagger_{\vec k}-v_k \hat b_{-\vec k},  \qquad  \hat a_{\vec k}= u_k \hat b_{\vec k}-v_k \hat b^\dagger_{-\vec k},
\end{equation}
where $\hat b^\dagger_{\vec k}$ and $\hat b_{\vec k}$ are operators of elementary excitations.
Thus, the Hamiltonian (\ref{he3}) reduces to the diagonal form $\hat H = E_0+\sum\limits_{\vec k} \varepsilon_k\hat b^\dagger_{\vec k}\hat b_{\vec k}$.
The Bogoliubov functions $ u_k,v_k$ are expressed in a standard way:
$ u_k,v_k=(\sqrt{\varepsilon_k/E_k}\pm\sqrt{E_k/\varepsilon_k})/2$ with $E_k=\hbar^2k^2/2m$ is the energy of free particle, 
and  the higher order Bogoliubov excitations energy is given by 
\begin{equation}\label {spec}
\varepsilon_k=\sqrt{[E_k- f(\vec k) n_c+\Sigma_{11}(\vec k)]^2-\Sigma_{12}(\vec k)^2},
\end{equation}
where  $\Sigma_{11}(\vec k)=2f(\vec k)n_c$ and $\Sigma_{12}(\vec k)=f(\vec k)(n_c+\tilde{m})$  are respectively, 
the first order normal and anomalous self-energies, $\tilde{m}$ is the anomalous density.\\
The spectrum (\ref {spec}) in principle cannot be used as it stands since it does not guarantee to give the best excitation frequencies 
due to the inclusion of the anomalous average which leads to the appearance of a gap in the excitation spectrum \cite {Burnet, Yuk, boudj2011}. 
One way to overcome this problem is to use the condition $\tilde{m}/n_c \ll 1$, which is valid at low temperature 
and necessary to ensure the diluteness of the system. Otherwise, the gas becomes strongly correlated and, thus, the Bogoliubov approach fails for $\tilde{m}/n_c \gg 1$.

Assuming now the limit $\tilde{m}/n_c \ll 1$, the normal and anomalous self energies simplify to $\Sigma_{12}(\theta_k)=\mu_0(\theta_k)$ and $\Sigma_{11} (\theta_k)=2\mu_0(\theta_k)$ 
where $\mu_0=n_c\lim\limits_{k\rightarrow 0} f({\vec k})$ is the chemical potential defined in the first order of perturbation theory \cite{Bel}. 
Therefore, the excitation frequency (\ref {spec}) reduces to
 \begin{equation}\label {spec1}
\varepsilon_k=\sqrt{E_k^2+ 2\mu_0(\theta_k) E_k},
\end{equation}
which is a gapless specrtum.\\
It is also easy to check that the HP theorem \cite{HP} $\Sigma_{11}(\theta_k) -\Sigma_{12}(\theta_k)=\mu_0(\theta_k)$, is well satisfied.\\
For $k\rightarrow 0$, the excitations are sound waves $\varepsilon_k=\hbar c_{sd} (\theta_k) k$, where $c_{sd} (\theta_k)=c_{s}\sqrt{1+\epsilon_{dd} (3\cos^2\theta_k-1)}$
with $c_{s}=\sqrt{gn_c/m}$ is the sound velocity without DDI.\\
Due to the anisotropy of the dipolar interaction, the self energies and the sound velocity acquire a dependence on the propagation direction, which is fixed by the angle
$\theta_k$ between the propagation direction and the dipolar orientation. This angular dependence of the sound velocity has been confirmed experimentally \cite {bism}.

The noncondensed and the anomalous densities are  defined as  $\tilde{n}=\sum_{\vec k} \langle\hat a^\dagger_{\vec k}\hat a_{\vec k}\rangle$ 
and $\tilde{m}=\sum_{\vec k} \langle\hat a_{\vec k}\hat a_{-\vec k}\rangle$, respectively. Then invoking for the operators $a_k$
the transformation (\ref {trans}), setting $\langle \hat b^\dagger_{\vec k}\hat b_{\vec k}\rangle=\delta_{\vec k' \vec k}N_k$ and putting the rest of the expectation values equal to zero,
where $N_k=[\exp(\varepsilon_k/T)-1]^{-1}$ are occupation numbers for the excitations.  
As we work in the thermodynamic limit, the sum over $\vec k$ can be replaced by the integral $\sum_{\vec k}=V\int d^3k/(2\pi)^3$ 
and using the fact that $2N (x)+1= \coth (x/2)$, we obtain:

\begin{eqnarray}\label {nor}
&&\tilde{n}=\frac{1}{2}\int \frac{d^3k} {(2\pi)^3} \left[\frac{E_k+\Sigma_{12}(\theta_k)} {\varepsilon_k}-1\right]\\&&+\frac{1}{2}\int \frac{d^3k} {(2\pi)^3} \frac{E_k+\Sigma_{12}(\theta_k)} {\varepsilon_k}\left[\coth\left(\frac{\varepsilon_k}{2T}\right)-1\right]\nonumber,
\end{eqnarray}
and
\begin{equation}\label {anom}
\tilde{m}=-\frac{1}{2}\int \frac{d^3k} {(2\pi)^3} \frac{\Sigma_{12}(\theta_k)} {\varepsilon_k}-\frac{1}{2}\int \frac{d^3k} {(2\pi)^3} \frac{\Sigma_{12}(\theta_k)} {\varepsilon_k}\coth\left(\frac{\varepsilon_k}{2T}\right).
\end{equation}
First terms in Eqs.(\ref{nor}) and (\ref {anom}) are the zero-temperature contribution to the noncondensed $\tilde{n}_0$ and anomalous $\tilde{m}_0$ densities, respectively.
Second terms represent the contribution of the so-called  thermal fluctuations and we denote them as  $\tilde{n}_T$ and $\tilde{m}_T$, respectively. 

Expressions (\ref{nor}) and (\ref {anom}) must satisfy the equality \cite {boudj2010, boudj2011, boudj2012}
\begin{equation}\label {heis}
I_k=(2\tilde{n}_k+1)^2-|2\tilde{m}_k|^2= \coth^2\left(\frac{\varepsilon_k}{2T}\right).
\end{equation}
Equation (\ref {heis}) clearly shows that $\tilde{m}$ is larger than $\tilde{n}$ at low temperature, so the omission of the anomalous density in this situation 
is principally unjustified approximation and wrong from the mathematical point of view.\\
The expression of $I$ allows us to calculate in a very useful way the dissipated heat $Q=(1/n)\int E_k I_k d^dk/(2\pi)^d$ 
for $d$-dimensional Bose gas\cite {Yuk, boudj2012}, where $n=n_c+\tilde{n}$ is the total density. 
Indeed, the dissipated heat or the superfluid fraction (see below) are defined through the dispersion of the total momentum operator of the whole system. 
This definition is valid for any system, including nonequilibrium and nonuniform systems of arbitrary statistics. In an equilibrium system, the average total momentum is zero. 
Hence, the corresponding heat becomes just the average total kinetic energy per particle. 

\section{Fluctuations at zero temperature}
\label {zeroT}
In this section we restrict ourselves to study the quantum fluctuations and their effects on the thermodynamics of the system. \\
Let us start by calculating the quantum depletion. At zero temperature ($\tilde{n}=\tilde{n}_0$),  the integral in Eq.(\ref{nor}) gives 
\begin{equation}\label {dep}
\frac{\tilde{n}}{ n_c}=\frac{ 8}{3} \sqrt{\frac{ n_c a^3}{\pi}} {\cal Q}_3(\epsilon_{dd}).
\end{equation}
The contribution of the DDI is expressed by the function ${\cal Q}_3(\epsilon_{dd})$, which is special case $j=3$ of
 ${\cal Q}_j(\epsilon_{dd})=(1-\epsilon_{dd})^{j/2} {}_2\!F_1\left(-\frac{j}{2},\frac{1}{2};\frac{3}{2};\frac{3\epsilon_{dd}}{\epsilon_{dd}-1}\right)$, where ${}_2\!F_1$ 
is the hypergeometric function. Note that functions ${\cal Q}_j(\epsilon_{dd})$ attain their maximal values for $\epsilon_{dd} \approx 1$ and become imaginary for $\epsilon_{dd}>1$.\\
Equation (\ref {dep}) is formally similar to the that obtained from the zeroth order of perturbation theory \cite {lime}. 
The density $n_c$ of condensed particles  which constitutes our corrections, appears as a key parameter instead of the total density $n$. 

Now if we use the integral in Eq.(\ref{anom}) directly by summing over all states, we find that the expression for $\tilde{m}$ diverges as we take the sum over higher and
higher states i.e. the so called ultraviolet divergence.  The price to be paid  to circumvent this divergence is to introduce the Beliaev-type second order coupling constant \cite {lime, peth}
\begin{equation}\label {cc}
g_R (\vec k)=f(\vec k) -\frac{m}{\hbar^2}\int \frac {d^3q}{(2\pi)^3} \frac{f(-\vec q) f(\vec q)}{2E_k}.
\end{equation}
After the subtraction of the ultraviolet divergent part, the renormalized anomalous density is given \cite {Griffin} 
\begin{equation}\label {MR}
\tilde{m}_R=-n_c \int \frac{d^3k} {(2\pi)^3} f(\vec k) \left[\frac{1} {2\varepsilon_k}\coth\left(\frac{\varepsilon_k}{2T}\right)- \frac{1} {2E_k} \right] .
\end{equation}
In contrast to $\tilde{m}$ in (\ref{anom}),  $\tilde{m}_R$ has no ultraviolet divergence from large $k$ contributions.
The authors of \cite {Burnet1} have pointed out that the self-consistent ladder diagram approximation for the $T$-matrix can be expressed in terms of  $\tilde{m}_R$.\\
To obtain an estimate value of  $\tilde{m}$, we note that the quasi-particle energy goes over to the free
particle energy for $\varepsilon_k>gn_c$. At zero temperature ($\tilde{m}=\tilde{m}_0$), we find
\begin{equation}\label {anom1}
\frac{\tilde{m}}{n_c}=8\sqrt{\frac{ n_c a^3}{ \pi}} {\cal Q}_3(\epsilon_{dd}).
\end{equation}
One should mention at this level that this expression has never been obtained before in the literature. \\
Equation (\ref {anom1}) is important in several respects: first of all, it shows that the anomalous density is three times larger than the noncondensed density 
whatever the type of the interaction. 
Second, $\tilde{m}$ has a positive value in argreement with the case of uniform Bose gas with pure contact interaction \cite{Yuk, boudj2012}.
Likewise, the anomalous density obtained in Eq.(\ref {anom1}) leads us to reproduce exactly the Lee-Huang-Yang (LHY) corrected equation of state \cite{LHY} (see below).  

Remarkably, we see from expressions (\ref {dep}) and (\ref {anom1}) that the noncondensed and the anomalous densities increase monotocally with $\epsilon_{dd}$. 
For a condensate with pure contact interactions (${\cal Q}_3(\epsilon_{dd}=0)=1$), $\tilde{n}$ and $\tilde{m}$ reduce to their usual expressions. 
While, for maximal value of DDI i.e. $\epsilon_{dd}\approx1$, they are 1.3 larger than their values of pure contact interactions 
which means that the DDI may enhance fluctuations of the condensate at zero temperature. 

The anomalous density manifests itself into the second-order correlation function as \cite {Glaub}
\begin{eqnarray}\label {correl}
&&G^{(2)}(r)=\langle\hat \psi^\dagger(r)\hat \psi^\dagger(r)\hat \psi(r) \hat\psi (r)\rangle \nonumber \\&&
 =n_{c}^2+\tilde{m}^2+2\tilde{n}^2+4\tilde{n}n_c+2\tilde{m}n_c.
\end{eqnarray}
Equation (\ref{correl}) is obtained using Wick’s theorem. Inserting then Eqs.(\ref {dep}) and (\ref {anom1}) into (\ref{correl}), we obtain
\begin{equation}\label {G}
\frac{G^{(2)}}{n^2}=1+\frac{64}{3}\sqrt{\frac{n_c a^3 }{\pi}} {\cal Q}_3(\epsilon_{dd}).
\end{equation}
This equation is accurate to the first order in  $\tilde{n}/n_c$ and $\tilde{m}/n_c$ and shows how the correlation function depends to the interaction parameter $\epsilon_{dd}$.

The presence of quantum fluctuations leads also to corrections of the chemical potential which are given by
$\delta \mu=\sum\limits_{\vec k} f(\vec k) [v_k(v_k-u_k)]=\sum\limits_{\vec k} f(\vec k) (\tilde{n}+\tilde{m})$ \cite{ Griffin, boudj2012, abdougora}. \\
Inserting the definitions (\ref{nor}) and (\ref {anom}) into the expression of $\delta \mu$, we find after integration:
\begin{equation}\label {chem}
\delta\mu=\frac{32}{3}gn_c\sqrt{\frac{n_c a^3}{\pi}} {\cal Q}_5(\epsilon_{dd}). 
\end{equation}
The total chemical potential is then written  as $\mu=\mu_0 (\theta_k)+\delta\mu$.
For $n_c\approx n$ and for a condensate with pure contact interaction (${\cal Q}_5(\epsilon_{dd}=0)=1$),
the obtained chemical potential excellently agrees with the famous LHY quantum corrected equation of state \cite{LHY}.

By integrating the chemical potential correction with respect to the density, one obtains beyond mean field the ground state energy as
\begin{equation}\label {energ}
E=E_0(\theta_k)+\frac{64}{15}Vgn_c^2\sqrt{\frac{n_c a^3}{\pi}} {\cal Q}_5(\epsilon_{dd}), 
\end{equation} 
where $E_0(\theta_k)=\mu_0(\theta_k) N_c /2$ with $N_c$ is the number of condensed particles.\\
Note that our formulas of the equation of state (\ref {chem}) and the ground state energy (\ref{energ}) constitute a natural extension of those obtained 
in Ref \cite {lime}.


At $T = 0$, the inverse compressibility is equal to $\kappa^{-1} = n^2\partial\mu/\partial n$. 
Then, using Eq.(\ref{chem}), we get 
\begin{equation}\label {compr}
\frac{\kappa^{-1}}{n^2}=\frac{\mu_0(\theta_k)}{n_c}+16g\sqrt{\frac{n_c a^3}{\pi}}{\cal Q}_5(\epsilon_{dd}). 
\end{equation}

One can also show that the shift of the sound velocity is $16g\sqrt{n_c a^3/\pi} {\cal Q}_5(\epsilon_{dd})$, which is consistent with the change in the compressibility $mc_s^2 = n \partial \mu/\partial n $ \cite {Lev} associated with the LHY correction in the equation of state (\ref {chem}).
Expanding the square root of the obtained formula with $\epsilon_{dd}=0$ in powers of the gas parameter $n_c a^3$, we recover easily the Beliaev
sound velocity of Bose gas with pure contact interaction $\delta c_s/c_{s}\approx 8\sqrt{n_c a^3/\pi}$ \cite {Bel, Lev}.

What is noticeable is that the chemical potential, the energy and the compressibility are increasing with dipole interaction parameter. 
For $\epsilon_{dd}\approx1$, these quantities are 2.6 larger than their values of pure contact interaction 
which means that DDI effects are more significant for thermodynamic quantities than for the condensate depletion and the anomalous density.

The Bogoliubov approach assumes that fluctuations should be small. We thus conclude from Eqs. (\ref {dep}) and (\ref {anom1}) that at $T = 0$, the validity of the 
Bogoliubov theory requires the inequality
\begin{equation}\label {Bog}
\sqrt{n_c a^3}  {\cal Q}_{3} (\epsilon_{dd})\ll 1.
\end{equation}
For $n_c=n$, this parameter differs only by the factor ${\cal Q}_3(\epsilon_{dd})$ from the universal small parameter of the theory, $\sqrt{na^3}\ll 1$, in the absence of DDI. 

\section{Fluctuations at finite temperature}
\label {finiteT}
We now generalize the above obtained results for the case of a spatially homogeneous dipolar
Bose-condensed gas at finite temperature. 

\begin{figure}
\centering
\includegraphics[scale=0.8, angle=0]{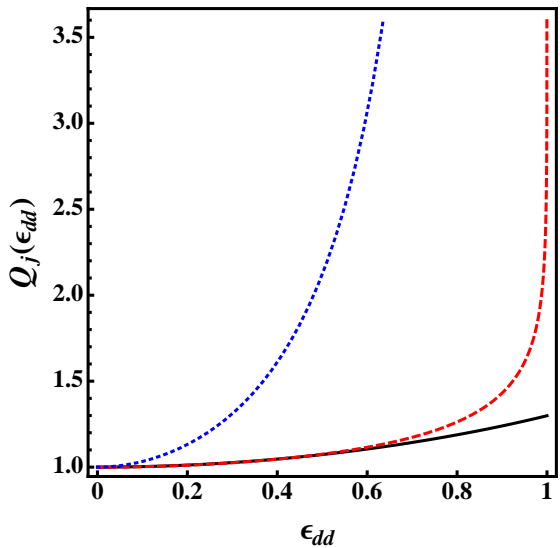}
\caption { Functions ${\cal Q}_{3}$ (solide line), ${\cal Q}_{-1}$ (red dashed line) and ${\cal Q}_{-5}$ (blue dotted line), which govern
the dependence of the condensate depletion, the anomalous fraction correction and superfluid fraction vs. the dipolar interaction parameter $\epsilon_{dd}$.}
\label{comp}
\end{figure}

At temperatures $T\ll gn_c $, the main contribution to integrals (\ref{nor}) and (\ref{anom}) comes from the region of small momentum where $\varepsilon_k=\hbar c_{sd} k$.
After some algebra, we obtain the following expressions for the thermal
contribution of the noncondensed and anomalous densities:

\begin{equation}\label {thfluc}
\frac{\tilde{n}_T}{n_c} =-\frac{\tilde{m}_T}{n_c} =\frac{2}{3}\sqrt{\frac{n_c a^3}{\pi}} \left(\frac{\pi T}{gn_c}\right)^{2}{\cal Q}_{-1}(\epsilon_{dd}). 
\end{equation}
Equation (\ref{thfluc}) shows clearly that $\tilde{n}$ and $\tilde{m}$ are of the same order of magnitude at low temperature and only their signs are opposite.  \\
Comparing the result of Eq. (\ref {thfluc}) with the zero-temperature noncondensed $\tilde{n}_0$ and anomalous $\tilde{m}_0$
densities following from Eqs. (\ref {dep}) and (\ref {anom1}) we see that at temperatures $T\ll gn_c$,
thermal contributions $\tilde{n}_T$ and $\tilde{m}_T$ are small and can be omitted when calculating the total fractions.
The situation is quite different at temperatures $T\gg gn_c$, where the main contribution to integrals (\ref{nor}) and (\ref{anom}) comes from the single particle excitations. 
Hence, $\tilde {n}_T\approx (mT/2\pi \hbar^2)^{3/2}  \zeta (3/2)$, where  $\zeta (3/2)$ is the Riemann Zeta function. 
The obtained $\tilde{n}_T$ is nothing else than the density of noncondensed atoms in ideal Bose gas. 
Moreover, the anomalous density being proportional to the condensed density, tend to zero together and hence their contribution becomes automatically small.\\
Another important remark is that for $\epsilon_{dd}\approx1$, thermal fluctuations (\ref{thfluc}) are 10.7 greater than their values of pure short range interaction.
This reflects that the DDIs may strongly enhance fluctuations of the condensate at finite temperature than at zero temperature (see figure.\ref{comp}). 

The same factor of Eq. (\ref {thfluc}) appears in the correction to the second order correlation function due to thermal fluctuations:
\begin{equation}\label {correlT}
\frac{G^{(2)}}{n^2}=\frac{8}{3}\sqrt{\frac{n_c a^3}{\pi}} \left(\frac{\pi T}{gn_c}\right)^{2}{\cal Q}_{-1}(\epsilon_{dd}). 
\end{equation}

Thermal fluctuations corrections to the chemical potential and the energy can be also obtained easily through expressions (\ref {thfluc}).


The Bogoliubov approach requires the conditions $\tilde{n}_T\ll n_c$ and $\tilde{m}_T\ll n_c$. 
Therefore, at temperatures $T \ll gn_c$, the small parameter of the theory turns out to be given as
\begin{equation}\label {BogT}
\frac{T}{ gn_c} \sqrt{n_c a^3}  {\cal Q}_{-1}(\epsilon_{dd})\ll 1.
\end{equation}
The appearance of the extra factor ($T/gn_c $) originates from the thermal fluctuations corrections. 

The superfluid fraction can be given as (c.f. \cite{LL9, abdougora, boudj2012})
\begin{equation}\label {supflui}
\frac{n_s}{n} =1-\int E_k\frac{\partial N_k}{\partial\varepsilon_k}\frac{d^3k}{(2\pi)^3}=1-\frac{2Q}{3T},
\end{equation}
where the quantity $2Q/3T$ represents the normal fraction of the Bose-condensed gas (liquid).\\
It is worth stressing that if in expression (\ref{supflui}) $\tilde {m}$ were omitted, 
then the related integral would be divergent leading to the meaningless value $n_s \rightarrow−\infty$.
This indicates that the presence of the anomalous density is crucial for the occurrence of the superfluidity in Bose gases \cite{boudj2013, Yuk1}
which is in fact understandable since both quantities are caused by atomic correlations.\\
Again at $T\ll gn_c$, a straightforward calculation leads to 
\begin{equation}\label {supflui1}
\frac{n_s}{n} =1-\frac{2\pi^2T^4}{45 mn\hbar^3 c_s^5} {\cal Q}_{-5}(\epsilon_{dd}).
\end{equation}
Remarkably, the normal density is $\propto T^4$, whereas the noncondensed density $\propto T^2$ as shown in (\ref {thfluc}).
One can see also from Eq.(\ref {supflui1}) that at $T=0$, the whole liquid is superfluid and $n_s=n$.
This shows that the normal density does not coincide with the noncondensed density $\tilde{n}$, 
and the superfluid density $n_s$ does not coincide with the condensed density $n_c$ of dipolar Bose gas.
At $T \gg gn_c$, there is copious evidence  that the normal density agrees with the noncondensed density of an ideal Bose gas.
Additionally, the normal density is rapidly increasing with the dipolar interaction as is depicted in figure.\ref{comp}. 

The system pressure can be expressed through $P=-(\partial F/\partial S)_T$, with the free energy given by $F=E_0+T \int\ln[1-\exp(-\varepsilon_k/T)] (d^3k/(2\pi)^3)$. 
At temperatures $T\ll gn_c$, the thermal pressure can be calculated as
\begin{equation}\label {therpress}
P_T=\frac{\pi^2T^4}{90 (\hbar c_s)^3} {\cal Q}_{-3}(\epsilon_{dd}).
\end{equation}

The inverse isothermal compressibility is proportional to $ (\partial P/\partial n)_T$
\begin{equation}\label {Isocomp}
\left(\frac{\partial P} {\partial n}\right)_T= -\frac{\pi^2T^4}{60 m\hbar^3 c_s^5} {\cal Q}_{-3}(\epsilon_{dd}) +\cdots ,
\end{equation}
where the zero temperature contribution to the compressibility is given by the expression (\ref{compr}). 


\section{Conclusion}
\label{conc}

In this paper, we have derived the first corrections to the elementary excitations of  homogeneous dipolar BEC
gases arising from effects of finite temperature perturbation theory (beyond mean field theory). 
Useful analytic expressions for the noncondensed and the anomalous densities are obtained. We find that these fluctuations are angular independence at zero and finite temperatures. 
We have shown that the anomalous density is larger than the noncondensed density at zero temperature while both quantities are comparable at $T\ll gn_c$. 
Our results show that the anomalous density changes its sign with increasing temperature in agreement with uniform Bose gas with pure
contact interaction \cite {Yuk, boudj2012}. It was also shown that the roton modes of trapped dipolar BEC (pancake geometry) 
serve to change the sign of the anomalous density near the trap center for largre values of $\epsilon_{dd}$\cite {Blak2}.
Indeed,  the importance of the anomalous density is ascribed rather to its modulus but not to its sign.
It is worth stressing that the qunatum depletion and the anomalous fraction are not yet observed experimentally and remain
challenging even for a condensate with pure contact interaction. 
Effects of dipolar interactions on quantum fluctuations and on thermodynamic quantities such as the chemical potential, the ground state energy and the compressibility 
are profoundly discussed. 
Although these effects are not considerable at zero temperature and there is almost no difference with the short-range interaction case, 
we believe that our results are important from the theoretical point of view since they clarify how 
the condensate fluctuations and thermodynamic quantities depend on the relative interaction strength on the one hand 
and they show how the anisotropy of the DDI involves these quantities on the other.
Moreover, we have pointed out that at finite temperature, the DDI may significantly enhance thermal fluctuations and the thermodynamics of the system 
supplying a real opportunity for a future experimental realization.
The validity criterion of the Bogoliubov approach is precisely determined at both zero and finite temperatures.

\section{Acknowledgements}
We would like to thank the LPTMS-France for a visit, during which part of this work was conceived.
Gora Shlyapnikov and Axel Pelster are acknowledged for valuable discussions and comments.

\end{document}